\documentstyle[12pt]{article}

\title{Quantum Maxwell-Bloch Equations for Spontaneous Emission 
       in Optical Semiconductor Devices}

\author{Ortwin Hess and Holger F. Hofmann\\ 
 Institute of Technical Physics, DLR\\
 Pfaffenwaldring 38-40, D--70569 Stuttgart, Germany}

\begin{document}

\maketitle

\begin{abstract}
We present quantum Maxwell-Bloch equations (QMBE) for spatially inhomogeneous
optical semiconductor devices taking into account 
the quantum noise effects which cause spontaneous emission 
and amplified emission. 
Analytical expressions derived from the QMBE are presented for the spontaneous 
emission factor $\beta$ and the far field pattern of amplified spontaneous emission
in broad area quantum well lasers.
\end{abstract}

\section{Introduction}

In an optical semiconductor device, spontaneously emitted light 
may have significant influence on the spatiotemporal dynamics of 
the (coherent) light field.
Undoubtedly, the spatial coherence of spontaneous emission and amplified spontaneous
emission is most important close to threshold or in devices
with a light field output dominated by spontaneous emission such as 
superluminescent diodes and ultra low threshold semiconductor lasers
\cite{Gun94,Yam93}. 
Even in more common devices such as VCSEL-arrays, gain guided or 
multi-stripe and broad area semiconductor lasers,
the spontaneous emission factor is modified by the
amplification and absorption of spontaneous emission into the non-lasing modes \cite{Hof98b}
and amplified spontaneous emission may be responsible for multi mode laser operation and 
for finite spatial coherence. 

In the present article, Quantum Maxwell-Bloch Equations (QMBE) are presented, which describe the
spatiotemporal dynamics of both, stimulated  amplification and 
(amplified) spontaneous emission on equal footing in terms of expectation values of
field-field correlations, dipole-field correlations, carrier densities,
fields and dipoles. In particular, the quantum dynamics of the interaction between
the light field and the carrier system is formulated in terms of Wigner 
distributions for the carriers and of spatially continuous amplitudes
for the light field. Analytical results for the spontaneous emission factor $\beta$
and the far field pattern of amplified spontaneous emission
in broad area quantum well lasers are discussed. 

\section{Quantum Maxwell-Bloch Equations}

The Quantum Maxwell-Bloch equations for the carrier density $N({\bf r}_\parallel)$
and the Wigner distributions of the 
field-field  ($I({\bf r}_\parallel;{\bf r^\prime_\parallel})$) and the 
field-dipole ($C({\bf r}_\parallel;{\bf r^\prime_\parallel},{\bf k_\parallel})$)
correlation
\begin{eqnarray}
\label{eq:qmbeN}
\lefteqn{\frac{\partial}{\partial t}  N\left({\bf r}_\parallel\right)=} 
\nonumber \\ & &
      - \gamma N\left({\bf r}_\parallel\right)
      + D_{amb} \Delta  N\left({\bf r}_\parallel\right)
      + j \left({\bf r}_\parallel\right) 
\nonumber \\ 
&&   +i g_0\sqrt{\nu_0} \; 
         \frac{1}{4\pi^2}\int d^2{\bf k}_\parallel 
\left( C\left({\bf r};{\bf r}_\parallel,{\bf k}_\parallel\right)_{z=0}
    -  C^*\left({\bf r};{\bf r}_\parallel,{\bf k}_\parallel\right)_{z=0}
\right)
\\ 
\label{eq:qmbeC}
\lefteqn{\frac{\partial}{\partial t} 
 C\left({\bf r};{\bf r^\prime}_\parallel,{\bf k}_\parallel\right) =}
\nonumber \\ & &
 - \left(\Gamma + i \Omega({\bf k}_\parallel)
                + i \frac{\omega_0}{2k_0^2}\Delta_{\bf r}\right)
    C\left({\bf r};{\bf r^\prime}_\parallel,{\bf k}_\parallel\right)
\nonumber\\ 
&&   +i g_0 \sqrt{\nu_0} \;
\left(  f^e_{eq}\left(k_\parallel; N({\bf r^\prime}_\parallel)\right)
      + f^h_{eq}\left(k_\parallel; N({\bf r^\prime}_\parallel)\right) - 1 \right) 
  I\left({\bf r};{\bf r^\prime}\right)_{z^\prime=0} 
\nonumber \\
 &&  +i g_0 \sqrt{\nu_0}\; 
  \delta({\bf r}_\parallel-{\bf r^\prime}_\parallel)\delta(z)\;
      f^e_{eq}\left(k_\parallel; N({\bf r}_\parallel)\right)  
      f^h_{eq}\left(k_\parallel; N({\bf r}_\parallel)\right)
\\ 
\label{eq:qmbeI}
\lefteqn{\frac{\partial}{\partial t} I\left({\bf r};{\bf r^\prime}\right) = }
\nonumber \\ & &
 - i\frac{\omega_0}{2k_0^2}
   \left(\Delta_{\bf r}-\Delta_{\bf r^\prime}\right)
         I\left({\bf r};{\bf r^\prime}\right)
\nonumber \\ 
&& -i g_0\sqrt{\nu_0} \;
  \frac{1}{4\pi^2}\int d^2{\bf k}_\parallel 
\left(
  C \left({\bf r};{\bf r^\prime}_\parallel,{\bf k}_\parallel\right)
 \delta(z^\prime) 
 -C^*\left({\bf r^\prime};{\bf r}_\parallel,{\bf k}_\parallel\right)
 \delta(z) \right).
\end{eqnarray}
describe the interaction of a quantum well with a quantized light field,
where the two-dimensional coordinates parallel to a quantum well at $z=0$
are marked by the index $\parallel$. 
In the QMBE, many particle renormalizations are not mentioned explicitly, but can 
be added in a straightforward manner \cite{Hes96}. 
Moreover, the Wigner distributions of electrons (e) and holes (h), 
have been approximated by corresponding quasi equilibrium  Fermi distributions
%
$
f_{eq}^{e,h}({\bf k}_\parallel, N({\bf r}_\parallel))
$
which are associated with a local carrier density of $N({\bf r}_\parallel)$ at a temperature $T$.
In Eq.~(\ref{eq:qmbeN}), $D_{amb}$ is the ambipolar diffusion constant, 
$j \left({\bf r}_\parallel\right)$ the
injection current density, $\gamma$ the rate of spontaneous 
recombinations by non-radiative processes and/or spontaneous emission into
modes not considered in $I\left({\bf r}_\parallel,{\bf r^\prime}_\parallel \right)$,
and $\Delta_{\bf r}$ is the Lapalcian with respect to ${\bf r}$. 
Another major feature of the semiconductor medium is the fact that 
the difference $\Omega({\bf k}_\parallel)$ between the dipole 
oscillation frequency and the band gap frequency $\omega_0$ is
a function of the semiconductor band structure. Here we model the
band structure by assuming parabolic bands such that 
$\Omega({\bf k}_\parallel)=\left(\hbar/(2 m^e_{eff})
                               +\hbar/(2 m^h_{eff})\right)
                          {\bf k}_\parallel^2$.
Moreover, the carrier dynamics of the interband-dipole 
$p\left({\bf r}_\parallel,{\bf k}_\parallel\right)$ and the
dipole part of the field-dipole correlation 
$C\left({\bf r_\parallel};{\bf r^\prime}_\parallel,{\bf k}_\parallel\right)$  
depend on a correlation of the electrons with the holes. The
relaxation of this correlation is modeled in terms of 
a rate of $\Gamma({\bf k})$ which may be interpreted as the total
momentum dependent scattering rate in the carrier system.

In the QMBE, the three dimensional intensity function of the light field
\mbox{$I({\bf r};{\bf r}^\prime)$}
may be interpreted as a single photon density matrix or as a 
spatial field-field correlation. 
The light-matter interaction is mediated by the correlation function
$C({\bf r};{\bf r}_\parallel^\prime,{\bf k}_\parallel)$ which describes the
correlation of the complex field amplitude at ${\bf r}$ and the complex
dipole amplitude of the ${\bf k}_\parallel$ transition
at ${\bf r}_\parallel^\prime$. The phase of this complex correlation 
corresponds to the phase difference between the dipole oscillations
and the field oscillations. 
Finally, the source of emission is the imaginary part of the field-dipole correlation
$C({\bf r};{\bf r}_\parallel^\prime,{\bf k}_\parallel)$. 
The quantum Maxwell-
Bloch equations thus show how this imaginary correlation originates either from
stimulation by $I({\bf r};{\bf r}^\prime)$ or spontaneously from 
the product of electron and hole densities.

\section{Spontaneous emission factor}

The spontaneous emission factor $\beta$ is generally 
defined as the fraction of spontaneous emission being emitted into the cavity mode
\cite{Ebe93}. 

On the basis of our theory, an analytical expression for $\beta$ 
may be obtained for
zero temperature where the k-space integrals may be solved analytically
on the assumption of $\Gamma$ 
being independent of ${\bf k}_\parallel$. 
The analytical result reads
\begin{equation}
\label{eq:beta}
\beta(\omega,\Omega_f) =
\frac{3\sigma}{2\pi\rho_{L}\; W\; L\; \Omega_f}
\left(\arctan \left(\frac{\Omega_f-\omega}{\Gamma+\kappa}\right)
+  \arctan \left(\frac{\omega}{\Gamma+\kappa}\right)\right),
\end{equation}
with the density of light field modes (at the band edge frequency $\omega_0$)
$\rho_{L} = \omega_0^2 \pi^{-2} c^{-3} \epsilon_r^{3/2} $.
The Fermi frequency $\Omega_f(N) = \pi\hbar\: M\: N$,
with the effective carrier mass
$M = \left(m^e_{eff}+m^h_{eff}\right) / \left( m^e_{eff} m^h_{eff} \right)$, 
expresses, in particular, the carrier density dependence of $\beta$. 
For $\Omega_f,\omega\ll \Gamma+\kappa$ we recover the result typically given in the
literature (e.g. \cite{Ebe93}) being independent of $N$. 
Fig.~\ref{beta} shows the deviation of the spontaneous emission factor from this value 
as the Fermi frequency $\Omega_f$ passes the point of resonance with the cavity mode.
Fig.~\ref{beta} illustrates, in particular, the carrier density dependence of 
$\beta$ for three modes with frequencies above the band gap frequency given by 
(a)~$\omega= 0$,
(b)~$\omega= 0.5 (\Gamma+\kappa)$, and
(c) $\omega= \Gamma+\kappa$. 
Most notably, $\beta$ is always smaller than the usual
estimate given by $\beta(\omega=\Omega_f=0)$, which
is based on the assumption of ideal resonance between
the transition frequency and the cavity mode.

\section{Farfield pattern of a broad area laser}

In large spatially inhomogeneous laser devices, the spatial coherence 
of amplified spontaneous emission defines an angular distribution
of the emitted light in the far field. Spatial coherence increases as
the laser threshold is crossed.
For $T=0$ we may obtain an analytical expression 
for the far-field intensity distribution 
of a broad area semiconductor laser \cite{PRA98qmbe}.
Figure \ref{farfield} shows the far field intensity distribution
for different carrier densities below threshold defined by a pinning 
carrier density $N_p$. 
Figure \ref{farfield}~(b) shows the intensity distribution for carrier
densities halfway towards threshold. Already, the intensity maxima move
to angles of $\pm 15^\circ$, corresponding to the  frequency at which 
the gain spectrum has its maximum. In the case of Fig.~\ref{farfield}(c), the
threshold region is very close to the pinning density. The peaks in the far
field pattern narrow as the laser intensity is increased.
Consequently the far field pattern indeed is a measure of the spatial coherence --
similar as the linewidth of the laser spectrum is a measure of temporal coherence.
It is therefore desirable to consider quantum noise effects in 
the spatial patterns of optical systems. In the context of squeezing, 
such patterns have been investigated by Lugiato and coworkers \cite{Lug92}. 
The laser patterns presented here are based on the same principles. 
Usually, however, the strong dissipation prevents squeezing in laser systems
unless the pump-noise fluctuations are suppressed \cite{Yam86}.

\section{CONCLUSIONS}

Quantum Maxwell-Bloch equations (QMBE) for 
spatially inhomogeneous optical  semiconductor devices have been presented, which
take into account the quantum mechanical nature of the light field as well as that of the
carrier system and thus describe the effects of coherent
spatiotemporal quantum fluctuations.
An example of the spatial coherence characteristics
described by the QMBE, an analytic expression for the 
density dependence of the spontaneous emission factor $\beta$
and spatial profiles of the far field distribution of a broad area edge 
emitting laser are discussed.

%

%
\begin{figure}[h]
\caption{Carrier density dependence of the spontaneous emission
factor $\beta$ for three modes with frequencies above the band gap
frequency given by 
(a)~$\omega= 0$,
(b)~$\omega= 0.5 (\Gamma+\kappa)$, and
(c) $\omega= \Gamma+\kappa$. 
$\beta_0=\beta(\omega=N=0)$ is determined by the geometry of the laser.
The carrier density is given in terms of the transition frequency at the
Fermi edge $\Omega_f$.}
\label{beta}
\end{figure}
\begin{figure}[h]
\caption{Far field intensity distributions for increasing 
values of the carrier density ${\cal N} = N/N_p$.
(a)~${\cal N} =  0.05, 0.1, 0.15$,
(b)~${\cal N} =  0.25, 0.5, 0.75$,
and 
(c)~${\cal N} =  0.90, 0.95, 0.99$.
The peaks appear at emission angles of $\pm 15^\circ$.
}
\label{farfield}
\end{figure}

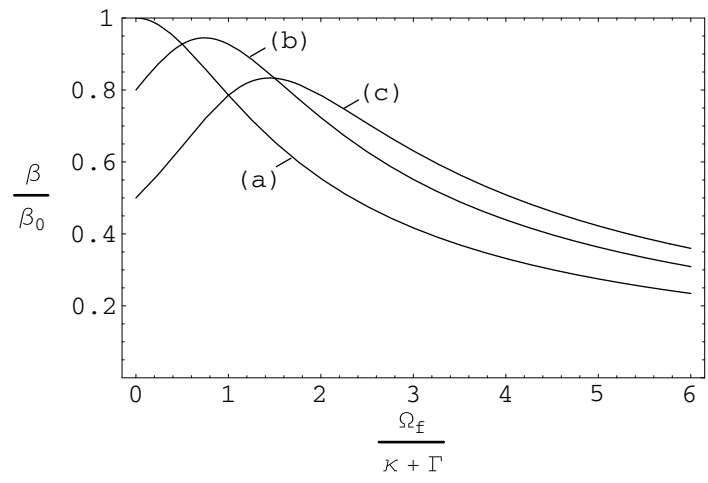
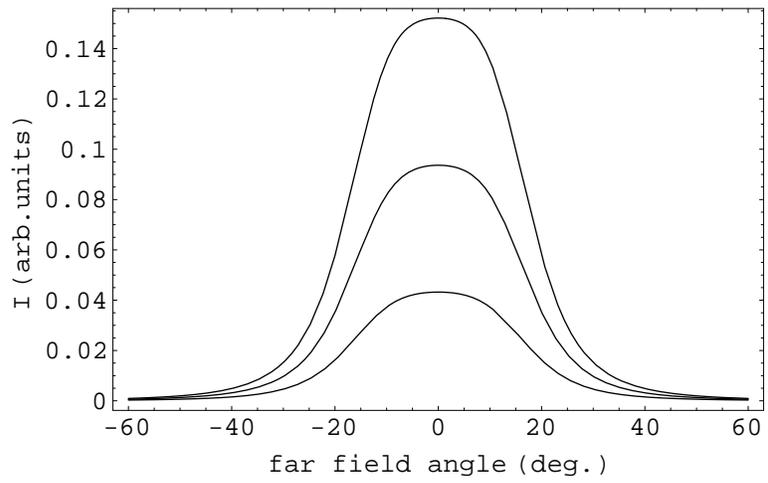
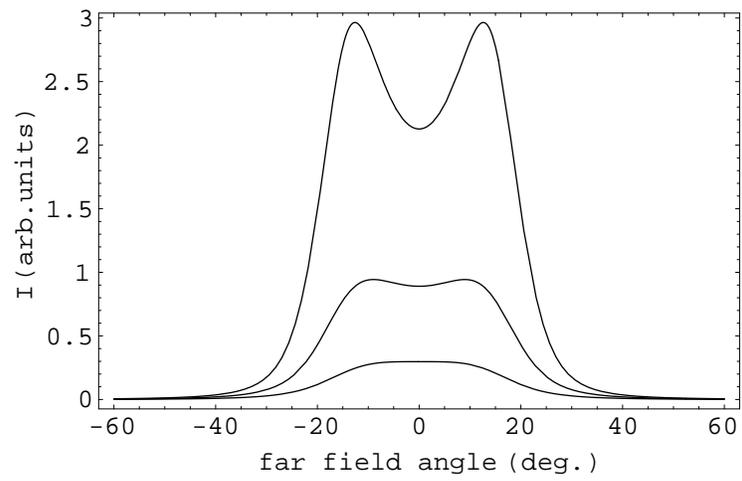
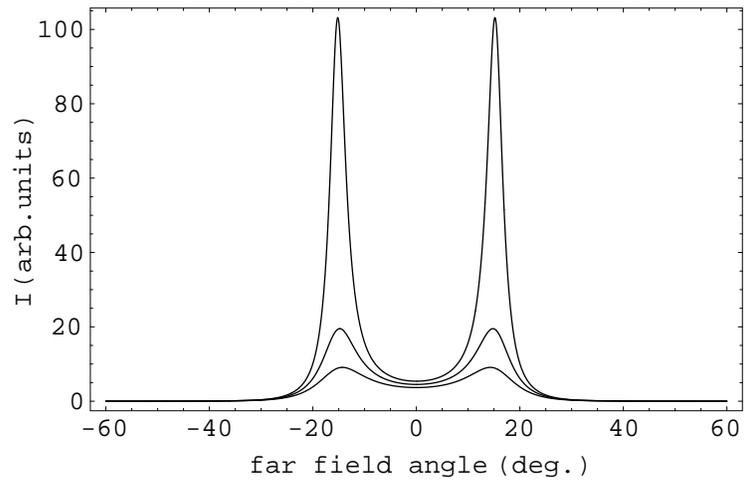
\end{document}